\documentclass[twocolumn,floatfix,superscriptaddress,amsmath,showpacs,showkeys,aps,prb]{revtex4-1}

\usepackage{t1enc}
\usepackage[final]{graphicx}
\usepackage{graphicx}
\usepackage{epsfig}
\usepackage{bm}
\usepackage[normalem]{ulem}
\usepackage[usenames]{color}
\usepackage{times}
\usepackage{verbatim}

\begin{document}

\title{Coarse grained models of stripe forming systems: phase diagrams,
anomalies and scaling hypothesis}

\author{Alejandro Mendoza-Coto and Daniel A. Stariolo}
\affiliation{Departamento de F\'\i sica, Universidade Federal do Rio Grande do Sul, CP 15051, 91501-970, Porto Alegre, Brazil}


\date{\today}

\begin{abstract}
Two coarse-grained models which capture some universal characteristics of 
stripe forming systems are studied. At high temperatures, the structure factors 
of both models attain
their maxima on a circle in reciprocal space, as a consequence of generic
isotropic competing interactions. Although this is known to lead to some
universal properties, we show that the phase
diagrams have important differences, which are a consequence of the particular
$\vec k$ dependence of the fluctuation spectrum in each model. 
The phase diagrams are computed in a mean
field approximation and also after inclusion of small fluctuations, which are shown
to modify drastically the mean field behavior. Observables like the modulation
length and magnetization profiles are computed for the whole temperature range
accessible to both models and some important differences in behavior are
observed. A stripe compression modulus is
computed, showing an anomalous behavior with temperature as
recently reported in related models. Also, a recently proposed scaling hypothesis
for modulated systems is tested and found to be valid for both models studied.
\end{abstract}




\maketitle

\section{Introduction}

Stripe forming systems are common in nature. 
Examples range from solid state systems, like ultrathin ferromagnetic films
~\cite{DeMaWh2000,PoVaPe2003}, stress-induced submonolayer islands on several 
substrates~\cite{MeLoAbBa2008} and strongly correlated electron systems~\cite{KiFrEm1998,
FrKi1999}, to soft matter systems like Langmuir monolayers~\cite{SeMoGoWo1991},
block copolymers~\cite{VeHaAnTrHuChRe2005,RuBoBl2008},
colloids and soft core systems~\cite{MaPe2004,ImRe2006,GlGrKaKoSaZi2007}. 
The origin of stripe structures can be traced back to some strong
universal mechanisms~\cite{EdJa2010}. At the heart of stripe formation there is
usually a competition between different kinds of interactions which lead to frustration
and the formation of patterns or microphases.
Common examples involve competition between a short range attractive interaction and
a long range repulsion~\cite{SeAn1995}. 

Although several general properties of stripe forming systems are well known, as
mean field phase diagrams and the behavior of the order parameter near the transition
temperature~\cite{Br1975,GaDo1982,BaRoFrGl1988}, the behavior at lower temperatures is less explored and new phenomena 
continue to appear. An example of this is the recently reported reentrance in the
phase diagram of frustrated dipolar ferromagnets in an external magnetic field
~\cite{SaLiPo2010,CaCaBiSt2011}. 

These facts motivate us to look for a deeper understanding of the phase behavior of
these kinds of systems. In the present work we address several questions which to a
large extent remain open. We present a complete study of two very well known models
of stripe forming systems, defined at a coarse-grain level, in which the main characteristic
defining the phase behavior of the systems is the form of the effective spectrum of fluctuations,
i.e. the wave vector dependence of the bare inverse correlation at low energies. This is
known to determine the existence and type of phase transition, at least at mean field level. 
We define a coarse-grained
free energy which is suitable for analyzing the low temperature behavior of relevant quantities,
like magnetizations and modulation lengths, besides allowing the determination of the
complete phase diagrams, in the whole temperature range. We show results from mean field
approximation and also when fluctuations are taken into account. An efficient numerical
implementation of the solutions allows us to compute several quantities with high accuracy.

The two models studied here
are usually considered in the literature as yielding essentially the same physics. Instead,
we show that besides obvious common characteristics, both models have very different phase
diagrams and very different behaviors of important quantities, like the temperature 
dependence of modulation length. By introducing fluctuations to the mean field results we
show that the nature of the transitions changes and the behavior of key quantities like
magnetizations are drastically affected by fluctuations. An important characteristic of
stripe forming systems with long range interactions is the temperature dependence of the
modulation length and of the domain wall widths. We compute the T dependence of the
modulation length in the whole T range, verify the well known square dependence near the transition
and show that our analytic results compare very well with recent experimental data~\cite{MeLoAbBa2008}
down to low temperatures.
We also introduce a measure of the width of domain walls and compare the T dependence of
the width relative the behavior near the transitions, where a single mode approximation usually
works fine. Finally, motivated by recent work on a scaling hypotesis for modulated 
systems~\cite{PoGoSaBiPeVi2010} we compute a response function, the stripe compressibility,
and show that it presents an anomalous behavior at low temperatures, in agreement with scaling
predictions of systems with power law competing interactions. Interestingly, the general scaling
behavior seems to work well also for a model which does not correspond to a system with long range
interactions, our model 2 defined below. This points to a wider applicability of the proposed
scaling hypotesis for stripe forming systems. 

The paper is organized as follows: in Sec. II we introduce the coarse-grain free energy in terms
of a generic spectrum of fluctuations. We define the equations of state at the mean field level
and when fluctuations are included and show how to solve them in an efficient way. In Sec. III
we introduce the two particular models studied, and present the main results on the phase
diagrams and temperature dependence of several quantities. In Sec. IV we present the results
of the stripe response function for both models. Finally, the conclusions of our work are
discussed in Sec. V .

\section{Mean Field approach including fluctuations}
\label{aa}

We consider a generic model in two spatial dimensions defined by the coarse-grained free energy:
\begin{widetext}
\begin{eqnarray}
\mathcal{H}[\phi]&=&\frac{\gamma}{2a^2}\int d^2 \vec{x}\,\left(\vec{\nabla}\phi \right)^2
+\frac{1}{2\delta a^2}\int d^2\vec{x}\int d^2\vec{x'}\ \phi(\vec{x}){J}(\vec{x}-\vec{x'})
\phi(\vec{x'}) +\nonumber \\
&  & \frac{1}{2\beta a^2}\int d^2\vec{x} \left\{ \left( 1+\frac{\phi(\vec{x})}{\phi_0} \right)\ln{\left( 1+\frac{\phi(\vec{x})}{\phi_0}\right)}
 +\left( 1-\frac{\phi(\vec{x})}{\phi_0} \right) \ln{\left(1-\frac{\phi(\vec{x})}{\phi_0}\right)}-2\ln{2} \right\}.
\label{Ham}
\end{eqnarray}
\end{widetext}
Here, $a$ represents the lattice spacing, $\beta=1/k_BT$ is the inverse temperature, $\gamma$ and $\delta$ are phenomenological 
constants and $\phi_0$ is the saturation value of the 
order parameter. The first term favors homogeneous configurations of the order parameter $\phi$ and the second term is intended to represent
a competing force, represented by the kernel $J(\vec x)$. These two terms can be considered the continuum limit of more microscopic
competing interactions, as e.g. exchange and dipolar interactions in thin magnetic films~\cite{GaDo1982}, hydrophobic and hydrophilic 
components in microemulsions or block copolymers~\cite{BaRoFrGl1988}, or charge separation due to long range Coulomb forces in low 
dimensional electron systems~\cite{FrKi1999}. The third term represents the mean field entropy. This effective Hamiltonian can be 
formally obtained. e.g. by a Hubbard-
Stratonovich transformation~\cite{BiDoFiNe1995} plus a coarse-grain process from an Ising model with competing short and long range interactions 
(represented by the non local term). 
A more common approach assumes an expansion of the entropy term for small $\phi$, which leads to the well known double well 
potential~\cite{GaDo1982,BiDoFiNe1995}. We will consider instead the
full form given by (\ref{Ham}), which allows to compute the phase diagrams down to low
temperatures.

Writing the quadratic part of (\ref{Ham}) in reciprocal space and normalizing the
order parameter field to its saturation value yields:
\begin{widetext}
\begin{equation}
\mathcal{H}[\phi] = \frac{\phi_0^2}{2a^2} \int d^2 \vec{k} \,A(\vec k)\,\phi_{\vec{k}}
\phi_{-\vec{k}}+
\frac{1}{2\beta a^2}\int d^2\vec{x}\, \left\{ \left(1+\phi(\vec{x})\right)\ln{\left(1+\phi(\vec{x})\right)}+\left(1-\phi(\vec{x})\right)\ln{\left(1-\phi(\vec{x})\right)}-2\ln{2}
\right\},
\label{Ham1}
\end{equation}
\end{widetext}
where the spectrum of fluctuations is given by
\begin{equation}
A(\vec k)= \gamma k^2+\frac{1}{2\delta}J(\vec k). 
\end{equation}

We consider systems in which the fluctuation spectrum has a single isotropic minimum 
at a non zero wave vector $k_0$. If $A(k_0)<0$ the system will develop modulated
structures. With some additional transformations it is possible to express the
effective Hamiltonian in dimensionless form:
 \begin{widetext}
\begin{equation}
\frac{\mathcal{H}[\phi]}{E_o} = \int d^2 \vec{k}\,{\hat{A}}(k)\, \phi_{\vec{k}}\phi_{-\vec{k}}+
 \frac{1}{\hat{\beta}}\int d^2\vec{x}\left\{ (1+\phi(\vec{x}))\ln(1+\phi(\vec{x}))+(1-\phi(\vec{x}))\ln(1-\phi(\vec{x}))-2\ln{2}\right\},
\label{Ham2}
\end{equation}
\end{widetext}
where 
\begin{eqnarray}
\nonumber
E_0&=&\frac{\phi_0^2|A(k_0)|k_0^2}{2a^2} \\ \nonumber
\hat{A}(k)&=&\frac{A(k k_0)}{|A(k_0)|} \\
\hat{\beta}&=&\phi_0^2|A(k_0)|k_0^4\beta. 
\end{eqnarray}
In this way all magnitudes but the characteristic energy $E_0$ in expression (\ref{Ham2}) are dimensionless . This expression was used for all the numerical computations in this work.

The mean field solutions of (\ref{Ham2}) are given by:
\begin{equation}
\left.  \frac{\delta H[\phi]}{\delta \phi(\vec{x})} \right|_{\phi=
                    \langle\phi\rangle} = 0
\label{Meanf}
\end{equation}

Fluctuations to the mean field solution can be introduced in the form
$\phi = \langle \phi \rangle + \psi$. The new stationary solutions 
are given by:
\begin{equation}
 \left\langle \frac{\delta H[\langle\phi\rangle+\psi]}{\delta \langle\phi(\vec{x})\rangle} \right\rangle=0,
\label{Meanfluct}
\end{equation}
where the average is relative to the mean field measure.



\subsection{Stripes solutions}
It is well known that, in the absence of external fields, the solutions which minimize 
the free energy of model (\ref{Ham2}) are one dimensional modulations of
the order parameter, i.e. stripe solutions in d=2, which can be written as:
\begin{equation}
\phi(x)=\sum_{i=0}^{\infty}m_i\sin\left(2\pi(2i+1)\frac{x}{\lambda}\right),
\label{phi}
\end{equation}
where $\lambda$ is the ``modulation length''.
In all calculations, the number of modes considered was $i_{max}=60$. 
This allow us to make a full characterization of the modulation profiles, 
as shown in the following sections. 

The mean field equation (\ref{Meanf}) reads:
\begin{equation}
\phi(x)=\tanh\left[-\hat{\beta}\int\frac{d^2\vec{k}}{(2\pi)^2}e^{\mathit{i}\vec{k}\cdot\vec{x}}\hat{A}(k)\phi_{\vec{k}} \right]
\label{phieq}
\end{equation}

To obtain the solution at a given temperature, equations (\ref{phi}) and
(\ref{phieq}) are solved for fixed $\lambda$. Then the free energy is minimized 
with respect to $\lambda$ to get the final solution.

\subsection{Stripes solutions including fluctuations}

Up to second order in the fluctuation fields, equation (\ref{Meanfluct}) is given by:
\begin{equation}
 \frac{\delta H[\langle\phi\rangle]}{\delta \langle\phi(\vec{x})\rangle}-\frac{1}{\hat{\beta}}\frac{2\langle\psi^2\rangle(\vec{x})\langle\phi\rangle(\vec{x})}{(1-\langle\phi\rangle^2(\vec{x}))^2}=0
\label{Meanf1}
\end{equation}    
In order to solve this equation the mean squared local fluctuations 
$\langle\psi^2\rangle(\vec{x})$ have to be computed. To do this 
we used the following method: let's consider the field $\phi(\vec x)$ in the 
presence of a local unknown molecular field due to the rest of the system.

The partition function for a single site in an external field $h$ is:
\begin{equation}
Z(h)=\int_{-1}^{1}d\phi \, \exp{[-\beta H_0(h,\phi)]}
\end{equation}
where
\begin{eqnarray}
\nonumber
{\beta} H_0(h,\phi)&=&-\beta h\phi +\frac{1}{2}\left[(1+\phi)\ln(1+\phi)\right.
              \\ \nonumber
&+ & \left. (1-\phi)\ln(1-\phi)-\ln{2}\right]
\end{eqnarray}

\begin{figure}
\begin{center}
\includegraphics[scale=0.5,angle=0]{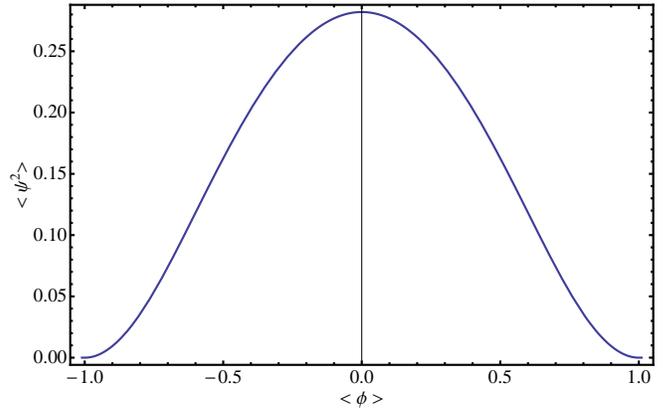}
\caption{(Color on-line) Local relation between the fluctuations amplitude and the mean order parameter. \label{1}}
\end{center}
\end{figure}

Then,
\begin{eqnarray}
\nonumber
\langle\phi\rangle(h)&=&\frac{1}{Z(h)}\int_{-1}^{1}d\phi \ \phi\exp(-\beta H_0(h,\phi))\\ \nonumber
\langle\phi^2\rangle(h)&=&\frac{1}{Z(h)}\int_{-1}^{1}d\phi \ \phi^2\exp(-\beta H_0(h,\phi))\\
\langle\psi^2\rangle(h)&=&\langle\phi^2\rangle(h)-(\langle\phi\rangle(h))^2
\label{parametric}
\end{eqnarray}
Of course, the value of the local molecular field $h$ remains unknown. 
Numerically, we can consider it as a free parameter to determine parametrically the non
trivial relation between the mean square fluctuations $\langle\psi^2\rangle(\phi)$
and the mean local order parameter. The parametric solution of the system 
(\ref{parametric}) is shown in Fig.\ref{1}. It is important to note that,
as a result of the temperature dependence of the potential used, the relation
 between $\langle\psi^2\rangle$ and $\langle\phi\rangle$ is temperature independent. 

The equation of state including the fluctuation fields now reads:
\begin{eqnarray}
\nonumber
\phi(x)&=&\tanh{\left[-\hat{\beta}\int\frac{d^2\vec{k}}{(2\pi)^2}e^{\mathit{i}\vec{k}\cdot\vec{x}}\hat{A}(k)\phi_{\vec{k}}\right.} \\
&+&\left. \frac{2\psi^2(\phi(x))\phi(x)}{(1-\phi^2(x))^2}\right].
\label{mfluc}
\end{eqnarray}

The effect of fluctuations is to add a new term in the effective
 field, which tends to rise up the order parameter profile in the region corresponding to domain walls. This means that when fluctuation effects are present 
profiles should be more sharp and correspondingly domain walls will be thinner. 

In the following we apply the method described above to solve equations
(\ref{phieq}) and (\ref{mfluc}) for two well known models with stripe phases.

\section{Phase diagrams.}
We focus our study in systems which present isotropic interactions, meaning that
$\hat{A}(\vec{k})=\hat{A}(k)$. Due to our 
selection of dimensionless variables, the minimum of the fluctuation spectrum is reached at $k=1$ and then $\hat{A}(1)=-1$ . Besides these universal characteristics, different
models are defined by the particular form of the fluctuation spectrum. The first 
model considered (model 1) is defined by~\cite{Br1975,BaSt2007-1,CaCaBiSt2011}:
\begin{equation}
\hat{A}(k)=-1+a(k-1)^2
\label{esp1}
\end{equation}
and the second (model 2) by~\cite{HoSw1995}:
\begin{equation}
\hat{A}(k)=-1+a(k^2-1)^2.
\label{esp2}
\end{equation}
Model 1 has a leading $\vec k$ dependence, for small $\vec k$, which is linear in $k \equiv
|\vec k|$, i.e. it is non-analytic. It is a good model of, e.g. an ultrathin ferromagnetic 
film with strong
perpendicular anisotropy, in which short range ferromagnetic
 exchange interaction competes with the long range dipolar interaction~
\cite{PoGoSaBiPeVi2010}. 
Model 2 is known as the Swift-Hohemberg model, and was introduced to describe the
physics at a convective instability~\cite{SwHo1977,HoSw1995}. Alternatively,
it can represent the continuum limit of a system with isotropic attractive nearest neighbors interactions plus repulsive next-nearest neighbor interactions~
\cite{TeSt1987,DaLiWi1988}.

Although both models are similar in that they have a single minimum at a non zero
wave vector modulus, they nevertheless show important differences which are shown
in what follows~\footnote{In a classic Landau approach the mean field transition is driven
by the most unstable mode $k_0$, which corresponds to the wave vector minimizing the spectrum. Very near the
transition, we can consider that $(k - k_0)$ is very small, and then it is easy to see that model two reduces to
model 1. Then the mean field transition yields essentially the same result. Nevertheless, as soon as the
temperature is reduced below the transition, other modes enter the scene and new phenomena emerges, as e.g. the
temperature dependence of the modulation lengths. Analytical and even numerical studies of the low temperature
phase are difficult and rare in the literature.}. To our knowledge a detailed comparison of models of this kind has
not been done up to now, and this frequently led to erroneous conclusions about
their behavior. 

\begin{subsection}{Model 1 }
 
Fig. \ref{2} shows the mean field phase diagram for model 1. The parameter 
$a$ is proportional to the curvature of the fluctuation spectrum at the minimum
wave vector:
\begin{equation}
 a=\frac{1}{2 |{A(k_0)}|} \left. \frac{d^2A(k)}{dk^2} \right|_{k=k_0}.
\end{equation}
Note that, due to our choice of dimensionless variables, the temperature dependence is included in $\hat{\beta}$,  defined as the inverse of $T/T_c$.
This is the reason why the critical line is horizontal. This line 
defines a second order phase transition between a low temperature modulated phase 
and  a high temperature disordered one. 
        
\begin{figure}
\begin{center}
\includegraphics[scale=0.5,angle=0]{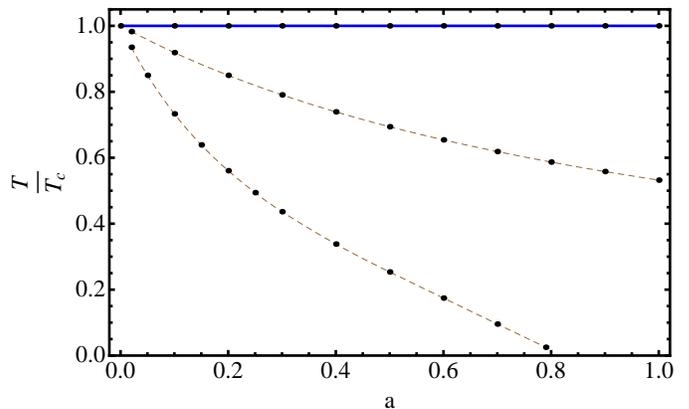}
\caption{(Color on-line) Mean field phase diagram for model 1. The dashed-dotted lines represent 
equal modulation length ($\lambda$) curves \label{2}.
 The upper full line (blue on-line) corresponds to a second order transition between a modulated and
a disordered phase. Dots represent the results of the numerical solution. }
\end{center}
\end{figure}

The dashed-dotted curves represent lines of constant modulation length $\lambda$. 
In fact, for fixed $a$ the modulation length decreases steadily as the temperature increases
towards $T_c$ and then curves nearer the critical line correspond to smaller values
of $\lambda$. The full temperature dependence of the
modulation length is shown in Fig. \ref{6} (dots), which shows that it varies
continuously with temperature with an asymptotic value corresponding to the
minimum of the spectrum. 
\begin{figure}
\begin{center}
\includegraphics[scale=0.51,angle=0]{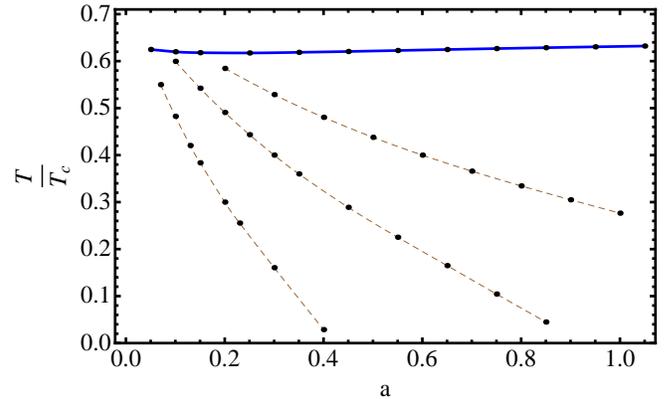}
\caption{(Color on-line) Mean field phase diagram for model 1 including fluctuations. The dashed-
dotted lines represent equal modulation length ($\lambda$) curves \label{3}.
 The upper full line (blue on-line) corresponds to a discontinuous transition between a modulated
and a disordered phase. Dots represent the results of the numerical solution.}
\end{center}
\end{figure}

The phase diagram of model 1 including fluctuations is shown in Fig. \ref{3}. At
first sight it seems similar to the mean field diagram. The critical temperature
is depressed relative to the mean field one, as expected when fluctuations are
included. The main difference with the mean field results is that the transition 
is discontinuous, i.e. 
fluctuations change the nature of the phase transition. In this case, the critical
line was determined at the temperatures where the difference between the free
energies of the disordered and modulated solutions changes sign. This characterizes
a first order transition induced by fluctuations.

The behavior of the lines of constant modulation length is different from that of the mean field case. In
Fig. \ref{2} all the lines of constant $\lambda$ approach a limiting value of
$\lambda=1$ at $T_c$, the value corresponding to the minimum of the spectrum $A(k)$,
as can be seen in Fig. \ref{6} (upper dots).
When fluctuations are included the transition line is not an equal modulation 
length curve, in such a way that the modulation length at the transition point is a 
decreasing function of $a$.

While it is well known that the modulation length grows quadratically with temperature
near the transition~\cite{PoVaPe2006,CaCaBiSt2011}, the low temperature behavior of $\lambda$
is less known. Based on scaling approaches from other
authors, Mentes et al.~\cite{MeLoAbBa2008}  proposed a numerical fit for their data on stripe forming ultrathin Pd/W(110) films. 
The proposed functionality, which takes into account 
fluctuation effects on the interface free energy and domain wall width, is the following:
\begin{equation}
 \lambda(T)=\lambda_0+a(1-bT)\exp(-cT).
\label{lambda}
\end{equation}
In Fig. \ref{6} we show the results obtained directly from our numerical solutions (dots)
together with the best fits from equation (\ref{lambda}) (solid lines) for both approximations, 
mean field and mean field plus fluctuations, showing that the physically motivated form (\ref{lambda})
is compatible with our numerical data.

\begin{figure}
\begin{center}
\includegraphics[scale=0.5,angle=0]{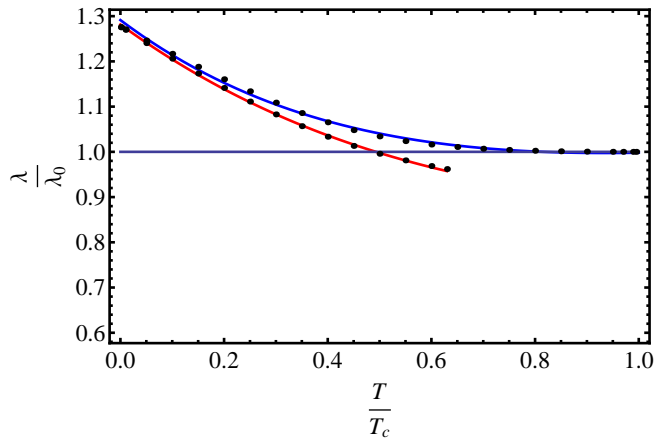}
\caption{(Color on-line) Temperature dependence of modulation length ($\lambda$) of model 1  
for $a=1$, normalized to the value corresponding to the minimum of the spectrum, 
$\lambda_0$. Superior and inferior dots correspond to the results of the case without fluctuations and 
 when fluctuations are included, respectively. The full lines represent the best fits of numerical 
results by the proposed functionality (\ref{lambda})\label{6}. The lower full line (red on-line) is 
interrupted because of the transition to the disordered phase.}
\end{center}
\end{figure}

In figure \ref{7} the behavior of the amplitude of the modulated solution is
shown. The temperature in the horizontal axis is scaled relative to the critical
temperature of the mean field solution. A notable fact is that the first order
transition induced by fluctuations is very strong, the amplitude stays at very
high values in the whole modulated phase, until the transition
takes place. This is also reflected in the behavior of the domain wall width
relative to the modulation length, which is shown in figure \ref{8}. 
The definition used for the domain wall width was 
\begin{equation}
\xi=2\, \frac{\int_0^{\frac{\lambda}{4}}f(x)x\ dx}{\int_0^{\frac{\lambda}{4}}f(x)\ dx}
\end{equation}

where the weight function $f(x)$ is $f(x)=M-\phi(x)$, and $M$ corresponds to the maximum value of the $\phi(x)$ profile. In our definition of $\phi(x)$ we fix the phase of the modulation, 
taking $\phi(0)=0$ and $\phi(x)$ positive in the interval $(0,\lambda)$. 
For example, for a sine like profile, the domain wall width is:
\begin{equation}
 \frac{\xi}{\lambda}=\frac{\int_0^{\frac{\pi}{2}}(1-\sin(\theta))\theta d\theta}{\pi\int_0^{\frac{\pi}{2}}(1-\sin(\theta)) d\theta}\approx0.1303
\end{equation}
This value would be a superior bound for the domain wall width, corresponding to the case of
single mode approximation.
In figures \ref{8} and \ref{14} this value was indicated by a horizontal line. We can see in each case how close profiles are of a sine function when the transition takes place.     

\begin{figure}
\begin{center}
\includegraphics[scale=0.5,angle=0]{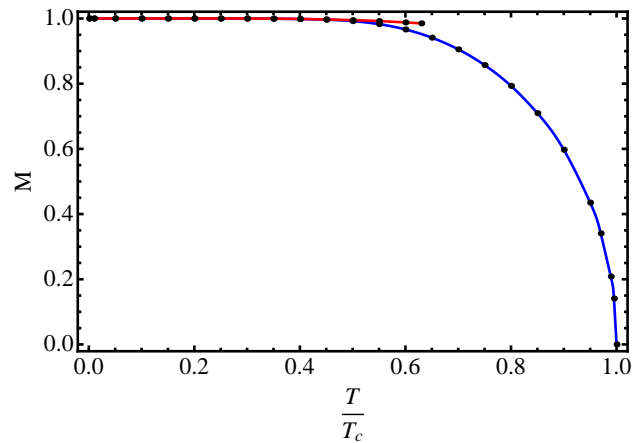}
\caption{(Color on-line) Temperature dependence of the modulation amplitude ($M$) for model 1
with $a=1$. The lower full line (blue on-line) corresponds to the mean field solution and the upper full one (red on-line) 
is the solution when fluctuations are included \label{7}. Dots represent the results of the numerical solution.}
\end{center}
\end{figure}

\begin{figure}
\begin{center}
\includegraphics[scale=0.5,angle=0]{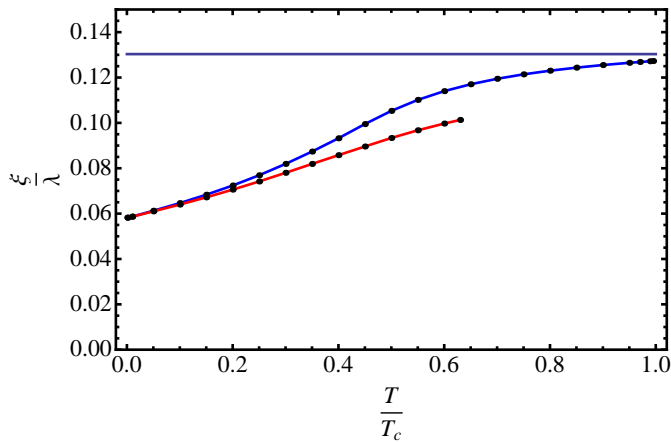}
\caption{(Color on-line) Temperature dependence of the domain wall width ($\xi$) normalized to the modulation length ($\lambda$) for model 1 with $a=1$.
 The superior full line (blue on-line) is the mean field solution and the lower one the solution with fluctuations included. The limit value of this magnitude 
for a perfect sine profile (single mode) is represented by the horizontal line. Dots represent the results of the numerical solution. \label{8} }
\end{center}
\end{figure}

\end{subsection}
\begin{subsection}{Model 2 }

The mean field phase diagram for model 2 is shown in Fig. \ref{4}. An important
difference with respect to the corresponding diagram of model 1 is the presence
of a uniform phase for small $T$ and $a$ values. This diagram has some similarity
with phase diagrams of anisotropic models with short range competing interactions,
like the ANNNI model~\cite{RaReTa2010}. Nevertheless, at variance with the ANNNI model,
in the present case there is no direct transition from the uniform to the 
disordered phases, at least up to very small values of $a$, where the precision
of our algorithm breaks down. Instead, a sequence of two phase transitions with
increasing temperature is observed. The modulated-uniform transition line 
characterizes a discontinuous transition at which the
modulation length diverges, as can be seen in figure \ref{20}. Note
that in the modulated region the free energy shows a single minimum at a
finite modulation length $\lambda$, but as the temperature is lowered a
second minimum at the far right of the figure appears. Numerically, this minimum
is compatible with an infinite value of $\lambda$, typical of a homogeneous
solution. 

\begin{figure}
\begin{center}
\includegraphics[scale=0.55,angle=0]{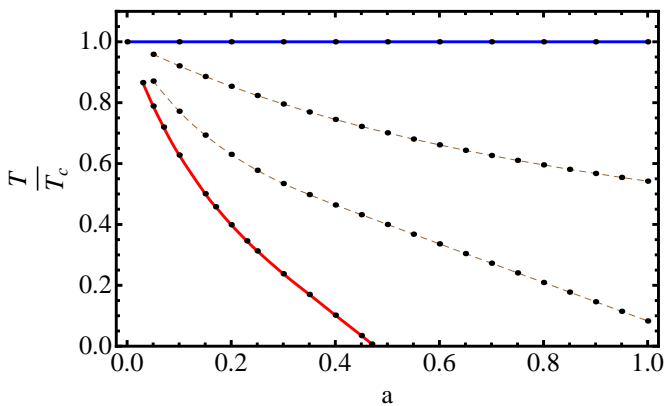}
\caption{(Color on-line) Mean field phase diagram for model 2. The dashed-dotted lines represent
 equal modulation length ($\lambda$) curves \label{4}.
 The full superior line (blue on-line) corresponds to a second order transition between a modulated and a
 disordered phase. The full lower line (red on-line) defines a transition between the modulated phase
and a uniform phase. Dots represents the results of the numerical solution. }
\end{center}
\end{figure}

\begin{figure}
\begin{center}
\includegraphics[scale=0.9,angle=0]{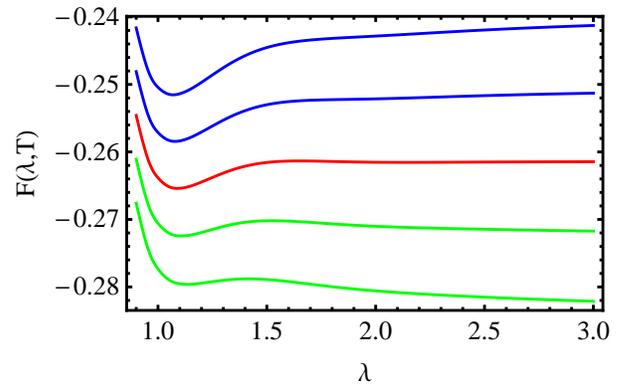}
\caption{(Color on-line) Free energy of model 2 as a function of modulation length for different temperatures.
Two top curves (blue on-line) show a single minimum at $\lambda$ slightly larger than 1.
In the two curves at the bottom (green on-line) a second minimum to the right of the figure
appears, signaling a discontinuous transition to a uniform state \label{20}. }
\end{center}
\end{figure}

Figure \ref{5} shows the phase diagram of model 2 when fluctuations to the mean
field solution are included. As was the case for model 1, a first effect of 
fluctuations is to depress the transition line between the modulated and disordered
lines. 

The curves of constant $\lambda$ for model 2 in the mean field approximation are shown in Figure \ref{4}. They tend smoothly to the critical line,
as in model 1 in such a way that higher curves corresponds to lower values of $\lambda$. The modulation length at the transition 
is a constant independent of $a$, corresponding to that of the minimum of the fluctuation spectrum. On the other hand when fluctuations are included
the shape of the equal $\lambda$ curves changes, as shown in  Figure \ref{5}.
 While upper curves still corresponds to lower values of the modulation length, in this case the modulation length at the transition is an
increasing function of $a$, at variance with the behavior of model 1, where the modulation length at the transition is a decreasing function 
of the parameter $a$.

\begin{figure}
\begin{center}
\includegraphics[scale=0.53,angle=0]{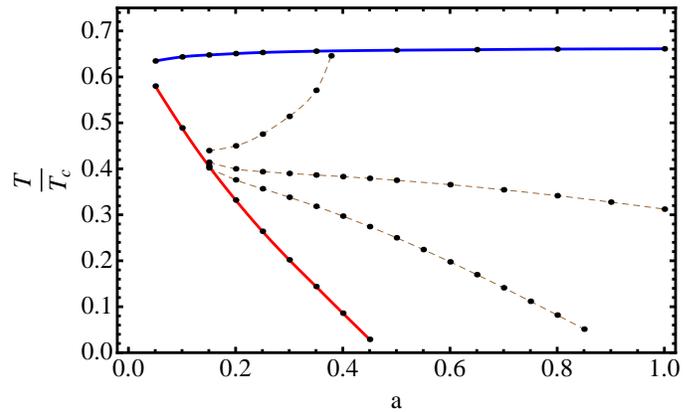}
\caption{(Color on-line) Mean field phase diagram for model 2 including fluctuations. The dashed-dotted lines represent equal modulation length ($\lambda$) curves \label{5}.
The top line (blue on-line) corresponds to a second order transition to a disordered phase. Below the full bottom line (red on-line) a uniform phase has less free energy 
than the modulated one. Note the bending of the constant $\lambda$ curves, meaning an increasing modulation length with $a$ at the transition temperature. Dots 
represent the results of the numerical solutions.}
\end{center}
\end{figure}

In figure \ref{12} the behavior of the modulation length with temperature is
shown for model 2. Note that the values of $\lambda$ stay very near $\lambda_0$,
the value at the transition, meaning that in the present model the modulation 
length has a weak dependence with temperature. This is an important difference 
with respect to the behavior of model 1. Nevertheless, as shown by the solid lines in
the same figure, a fit to the expression (\ref{lambda}) still works very well.

\begin{figure}
\begin{center}
\includegraphics[scale=0.45,angle=0]{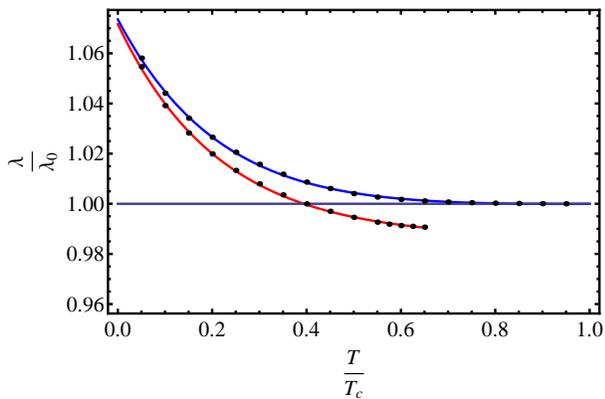}
\caption{(Color on-line) Temperature dependence of modulation length ($\lambda$) for model 2 with $a=0.5$. Superior and inferior dots correspond 
to the results without fluctuations and when fluctuations are included, respectively.
The full lines represent the best fits of numerical 
results by the proposed functionality (\ref{lambda}). The lower line (red on-line) is 
interrupted because of the transition to the disordered phase.\label{12}.}
\end{center}
\end{figure}

The behavior of the domain wall width for model 2 is shown in figure \ref{14}.
Comparing with figure \ref{8} for model 1,  we note that in the present model the profile is always 
very close to a sine, even at low temperatures. Besides that, fluctuations seem to be unimportant for the 
behavior of the domain wall width, at variance with the behavior observed in model 1. 
Nevertheless, the amplitude of the order parameter behaves 
similarly to that of model 1, staying practically at saturation until the
transition temperature is reached, in the case when fluctuations are included.
The fact that for model 2
the profiles stay very near the single mode case can be understood due to the presence of the quartic contribution in the 
fluctuation spectrum which implies that the energy required to excite higher harmonics 
at low temperatures is larger than in the quadratic case of model 1.

\begin{figure}
\begin{center}
\includegraphics[scale=0.55,angle=0]{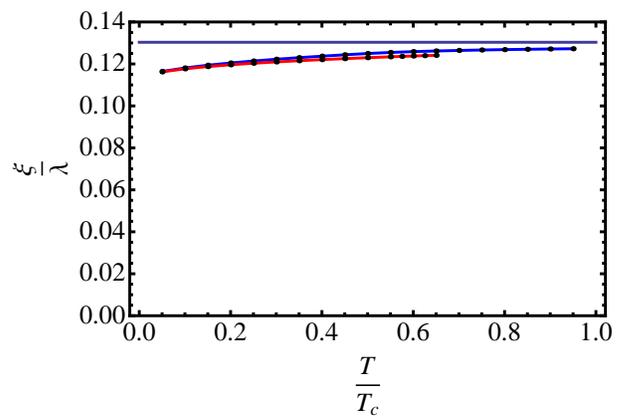}
\caption{(Color on-line) Temperature dependence of the domain wall width ($\xi$) normalized to the modulation length ($\lambda$) for $a=0.5$ in model 2. 
The upper line (blue on-line) is the mean field solution and the lower one (red on-line) the solution with fluctuations included. 
The limit value of this magnitude for a perfect sine profile (single
mode) is represented by the horizontal line. Dots represent the results of the numerical solution.\label{14}}
\end{center}
\end{figure}
\end{subsection}

\section{Anomalous behavior of stripe compressibility}
\label{anom}
A measure of a stripe compressibility in modulated systems can be obtained through the compression or Young modulus $B(T)$. 
This magnitude is related with the energy 
cost associated with deviations from the equilibrium modulation length ($\lambda(T)$) and it is defined as \cite{PoGoSaBiPeVi2010}:
\begin{equation}
 B(T)=\lambda(T)^2 \left. \frac{\partial^2F(\lambda,T)}{\partial\lambda^2}\right|_{\lambda(T)}
\label{B}
\end{equation}
We calculated the compression modulus directly from the free energy $F(\lambda,T)$. In a recent work, Portmann et al. \cite{PoGoSaBiPeVi2010}
have proposed some scaling relations for modulated matter. A scaling relation for the compression modulus was obtained in terms of the 
equilibrium modulation length ($\lambda(T)$) and the modulation amplitude ($M(T)$):
\begin{equation}
B=c\,M(T)^2\, \lambda(T)^{\Delta} ,
\label{esc}
\end{equation}
where $c$ is a constant and $\Delta$ is an exponent related with the system dimensionality and the microscopic nature of interactions. 
One of the main predictions of the proposed scaling relations is an anomalous behavior of the Young modulus when the exponent $\Delta$ is negative. This
is the case, e.g. in a dipolar frustrated ferromagnet in two dimensions, which has an effective free energy of the type of model 1 in
this work. More interestingly, Portmann and collaborators find the same form of scaling for the behavior of the critical magnetic field
as a function of the modulation length \cite{PoGoSaBiPeVi2010}. This implies a reentrant or inverse transition in the external field vs
temperature phase diagram at low temperatures for
all systems which have a negative value of the exponent $\Delta$. This interesting result is supported by experimental measures in 
ultrathin films of Fe/Cu(001) from the same group~\cite{SaLiPo2010}, and it has been reported in a Ginzburg-Landau model for the same 
system~\cite{CaCaBiSt2011}.

\begin{figure}
\begin{center}
\includegraphics[scale=0.75,angle=0]{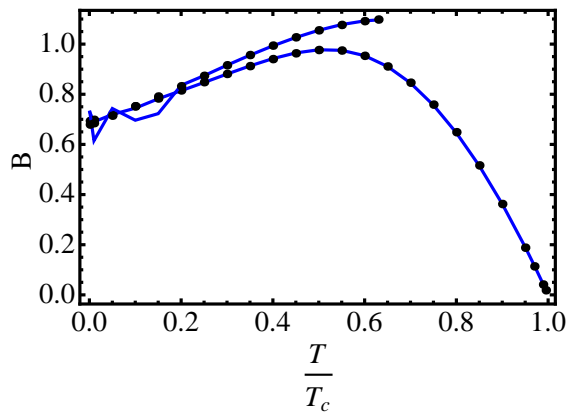}
\caption{(Color on-line) Continuous curves corresponds to the calculated Young modulus with and without fluctuations for model 1. The dots represent the best fit 
using the scaling relation equation (\ref{esc}). \label{16}}
\end{center}
\end{figure}

\begin{figure}
\begin{center}
\includegraphics[scale=0.88,angle=0]{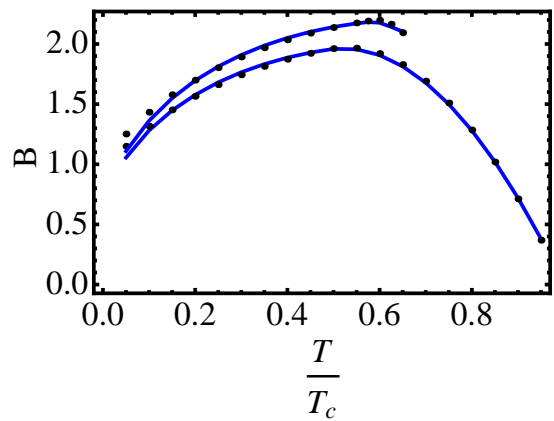}
\caption{(Color on-line) Continuous curves corresponds to the calculated Young modulus with and without fluctuations for model 2. The dots represent the best fit 
using the scaling relation equation (\ref{esc}). \label{17}}
\end{center}
\end{figure}

In Figures \ref{16} and \ref{17} we show the results of a direct calculation of the compression modulus (full curves) for both model 1 and model 2 
via equation (\ref{B}). 
The typical behavior of $B(T)$ in a normal system is a monotonically decreasing function of temperature, growing temperature leading to
weaker bonds between particles and a weaker elastic response. However we can see in Figures \ref{16} and 
\ref{17} a regime for low enough temperatures where the compression modulus is an increasing function of temperature. Furthermore, the inclusion 
of fluctuations in both models do not destroy this behavior but strengthen  the anomaly even more, which suggests a robust physics behind 
the numerical calculations. 
It is worth to mention that for very low temperatures the domain walls for model 1 become very sharp and as result numerical instabilities 
appear which can be seen in Figure \ref{16}.

We also tested the proposed scaling relation by fitting our analytical
results with expression (\ref{esc}). Results are shown with dots in Figures \ref{16} and 
\ref{17}.
In all cases the proposed scaling law fits very  well our numerical results.
As part of our checking of the scaling predictions we let the initially fixed (quadratic) exponent of $M(T)$  
in equation (\ref{esc}) to vary. The results for this exponent as well as for $\Delta$ 
in model 1 were very close to those predicted in \cite{PoGoSaBiPeVi2010} for the 2d dipolar frustrated ferromagnet. On the other hand, 
for model 2 important deviations where found. Nevertheless, in this case deviations for the values of exponents are
expected because, strictly speaking, model 2  does not correspond to the kind of systems considered by Portmann and collaborators.

It is interesting that the scaling relations seem to be valid for a wider class of models. Future work should address the extension
of the scaling relations to more general microscopic interactions as well as the origin
of the anomalous behavior of the Young modulus in stripe forming systems.

\section{Conclusions}
\label{conc}

We have presented several new results regarding two well known effective models for stripe 
forming systems in two dimensions. The models can be thought as the continuous limits of
some microscopic models with competing isotropic interactions, or alternatively, they can
be defined directly through the form of the high temperature structure factor. A common 
feature of both models is that the structure factor has its maximum in a ring-like region
in reciprocal space, which has strong consequences for the low temperature physics. 
Nevertheless, contrary to what is usually assumed in the literature, important differences
were found in their properties and phase diagrams that lie in the specific wave vector dependende of the high-T structure
factor, with Model 1 defined by a non-analytic dependence on the modulus of wave vector and
Model 2 defined by an analytic form. Model 1 is a good small $k$ representation of some
systems with long range interactions, like the dipolar frustrated ferromagnet and
stripe forming systems with elastic interactions with the subtrate, while Model 2
can be obtained as an effective model for competing finite range isotropic interactions, which leads
to an analytic expression for the $k$ dependence of the structure factor. 

We have computed the full phase diagrams of both models in two approximations: mean field, and 
mean field plus fluctuations. Relevant equations were solved numerically. In both cases it is observed
that fluctuations have a strong effect on the phase diagrams and other observables. With regard to
phase diagrams, the main effect of fluctuations is to change the order of the 
isotropic to modulated phase transition 
 from continuous in the mean field approximation to a
strongly discontinuous transition when fluctuations are taken into account. Besides the isotropic-modulated transition, Model 2 shows another discontinuous phase transition from the modulated to
a uniform (infinite modulation length) phase below a critical value of the
parameter $a$, which characterizes the curvature of the fluctuation spectrum at the minimum.

In both models, the modulation length $\lambda$ is a smooth function of temperature, characterizing
incommensurate modulations. The temperature dependence of $\lambda$ is strong for Model 1 and
weak for Model 2. While in the mean field approximation the modulation length attains its mimimum
value at the transition, which corresponds to the wave vector at the minimum of the structure
factor, when fluctuations are taken into account $\lambda$ can take values less than that
corresponding to the minimum of the structure factor. Another quantity of interest is the width
of the domain walls $\xi$. We have introduced an operational definition of the width, taking as a
reference value that corresponds to a single mode, sine-like solution, valid near the transition
in the mean field approximation. As was the case with the modulation length, $\xi$ has a stronger
dependence with temperature for Model 1. It is also important the fact that $\xi$ is smaller when
fluctuations are included, meaning sharper domain walls, in line with the sharp phase transitions
obtained in this case, where the amplitude of the modulations stay near saturation already below
the transition. 

Finally we have shown results for a response function, the compression modulus of stripes $B(T)$.
This quantity shows an anomaly in its behavior, when compared with normal systems, having a
maximum at intermediate temperatures. The anomalous behavior follows a recent prediction from a
scaling hypothesis for modulated systems, which was obtained in the context of models with some
particular long range interactions. We have verified its validity for both Model 1 and Model 2.
Interestingly, although Model 2 does not belong to the kind of models for which the scaling
hypothesis was predicted, it shows the same kind of scaling, with a particular scaling exponent.
The scaling exponent for Model 1 corresponds, to a good numerical approximation, to the value
predicted for a model with dipolar competing interactions.

In summary, we have computed some relevant quantities for models with modulated phases in two
dimensions, complementing old results and clarifiying some differences between two models
usually presented in the literature as giving essentially the same physics. We have shown that,
despite the obvious similarities, both models have important differences. An interesting outcome
of our results is the evidence of anomalous behavior in a response function in both models
considered, in line with recent results in related models and experiments in ultrathin magnetic
films. Future work should address the mechanisms that originate the anomalous behavior and the 
generality of this for other systems. Last but
not least, the straightforward numerical implementation of the coarse-grained models defined in
this work allowed us to access the whole temperature range of the phase diagrams, complementing
already known results, usually limited to temperatures near the phase transition lines.

\begin{acknowledgments}
We gratefully acknowledge financial support from CNPq (Brazil).
\end{acknowledgments}


%

\end{document}